# Understanding the anomalous thermoelectric behaviour of Fe-V-W-Al based thin films

Kavita Yadav[1]*, Yuya Tanaka[1], Kotaro Hirose[2], Masahiro Adachi[2], Masaharu Matsunami[1,3,4], Tsunehiro Takeuchi[1,3,4]

[1]Toyota Technological Institute, Hisakata 2-12-1, Tempaku, Nagoya 468-8511, Japan.

[2]Sumitomo Electric Industries Ltd., Transmission Devices Laboratory, 1-1-1, Koyakita, Itami-Shi, Hyogo 664-0016, Japan.

[3]Research Center for Smart Energy Technology, Toyota Technological Institute, Nagoya 468-8511, Japan

[4]CREST, Japan Science and Technology Agency, Tokyo 102-0076, Japan.

**ABSTRACT**

We have investigated the thermoelectric and thermal behaviour of Fe-V-W-Al based thin films prepared using radio frequency magnetron sputtering technique at different base pressures (0.1 ~ 1.0 X $10^{-2}$ Pa) and on different substrates (*n*, *p* and undoped Si). Interestingly, at lower base pressure, formation of bcc type of Heusler structure was observed in deposited samples, whereas at higher base pressure, we have noted the development of non-Heusler amorphous structure in these samples. Our findings indicates that the moderately oxidized Fe-V-W-Al amorphous thin film deposited on *n*-Si substrate, possesses large magnitude of $S \sim -1098 \pm 100$ μV$K^{-1}$ near room temperature, which is almost the double the previously reported value for thin films. Additionally, the power factor indicated enormously large values ~ 33.9 mW$m^{-1}K^{-2}$ near 320 K. The thermal conductivity of the amorphous thin film is also found to be 2.75 W$m^{-1}K^{-1}$, which is quite lower compared to bulk alloys. As a result, the maximum figure of merit is estimated to be extremely high i.e. ~ 3.9 near 320 K, which is among one of the highest reported values so far. The anomalously large value of Seebeck coefficient and power factor has been ascribed to formation of amorphous structure and composite effect of thin film and substrate.

## I. INTRODUCTION

In past few decades, thermoelectric (TE) devices have attracted considerable interest due to their capability of converting heat into electricity using Seebeck effect [1]. The generated power from these devices mostly depends on the temperature difference between the hot ($T_h$) and the cold ($T_c$) sides of the device. The performance of the thermoelectric generators is generally described in terms of thermoelectric figure of merit ($ZT = S^2 \sigma T / \kappa$) of their component thermoelectric materials. Here $S$, $\sigma$, $\kappa$, and $T$ are the Seebeck coefficient, electrical conductivity, total thermal conductivity, and absolute temperature at which the thermoelectric properties is measured, respectively [2-6]. In order to enhance the thermoelectric performance of any material, the power factor $PF = S^2\sigma$ in the numerator of $ZT$ should be improved and total thermal conductivity $\kappa = \kappa_{el} + \kappa_{latt}$ in the denominator must be decreased ($\kappa_{el}$ and $\kappa_{latt}$ represents the electronic and lattice contributions to $\kappa$) [7]. It is important to note here that these two parameters are dependent on each other. Thus, improving one parameter without affecting the other factor is rather difficult. Amongst them, $\kappa_{latt}$ is the only parameter which can be altered without substantially affecting the other parameters. In order to reduce $\kappa_{latt}$, different effective methods have been adopted such as nano-structuring, heavy element partial substitution, or enhancement in the phonon scattering events [8, 9]. Alternative ways to achieve high $ZT$ is reducing the dimensions or dimensionalities of the material under investigation.

In the present scenario, the commercially available thermoelectric modules working near room temperature are based on Bi-Te [10, 11]. However, the persistent issue with these materials is highly toxic and expensive nature of tellurium, thus, making it unfeasible to be used for thermoelectric applications. Hence, the demand to develop new alternative TE materials with high TE performance has been increasing in scientific community.

In this context, $Fe_2VAl$ based Heusler alloys can be one of the potential candidates for such type of thermoelectric applications. This alloy is solely composed of harmless, highly abundant, and inexpensive elements. Additionally, it displays a high-$PF$ as large as $Bi_2Te_3$ (~ 5 mWm$^{-1}$K$^{-2}$) near room temperature [12-15]. However, the high thermal conductivity of $Fe_2VAl$ in its pristine form ($\kappa$ ~ 29 W m$^{-1}$K$^{-1}$) makes the value of $ZT$ smaller, and therefore $Fe_2VAl$ based thermoelectric materials have been considered as undesirable ones for thermoelectric applications [16]. Heavier element substitutions (like Ta, W,) and micro- and nano- grain structuring have been extensively done to enhance the figure of merit in both bulk and thin films. The developed micro grained $Fe_2VAl$ and $Fe_2VAl_{0.95}Ta_{0.05}$ were reported to exhibit $ZT \sim 0.2$ near 400 K [17]. Mikami *et al.* have also obtained a relatively high $ZT \sim 0.20$ near 400 K in $Fe_2V_{0.9}W_{0.1}Al$ [18]. For the same composition, Hinterleitner *et al.* observed a slightly higher value of $ZT \sim 0.22$ [19]. These values were further enhanced to $ZT \sim 0.3$ in micro-grained $Fe_2V_{0.95}Ta_{0.05}Al_{0.9}Si_{0.1}$ [20] and in nano grained $Fe_2VAl_{0.95}Ta_{0.05}$ [21].

Recently, Hinterleitner *et al.* reported an exceptionally large $ZT \sim 6$ near 380 K in a $Fe_2V_{0.8}W_{0.2}Al$ thin film [22]. This value is found be approximately 30 times higher than the bulk $Fe_2V_{0.9}W_{0.1}Al$ synthesized by same group [19]. They have attributed the high magnitude of $ZT$ to (i) formation of bcc-W type of crystal structure in $Fe_2V_{0.8}W_{0.2}Al$ and (ii) the presence of large change in electronic density of states near the chemical potential, which was found to be in analogy with observed large value of Seebeck coefficient ($\sim -550$ µV/K). This exceptional finding led to curiosity among the researchers to explain and reproduce such high value of $ZT$ in similar materials.

Apart from experimental studies, various theoretical calculations have been carried out to understand the factors responsible for such a high thermoelectric performance in these thin films. Recently, Matsuura *et al.* [23] have reported that magnon drag mechanism can be responsible for high value of $ZT$. On the other hand, computational studies done by Alleno *et al.* has ruled out the possibility of high $ZT$ in $Fe_2V_{0.9}W_{0.1}Al$ based thin films [24]. Consequently, research to improve the $ZT$ and primarily the power factor of Fe–V–Al films needs to be pursued. Various external factors such as base pressure (B.P) of the sputtering system, and nature of substrate, which can significantly influence the nature of thermoelectric behaviour of these films have not been well explored. Hence, with this motivation, we explored, in this study, the effect of base pressure and nature of substrate on the thermoelectric properties of Fe-V-W-Al thin films.

In this manuscript, we report a detailed investigation of the electronic and thermal properties of Fe-V-W-Al based thin films. The thin films are prepared by means of radio frequency magnetron sputtering in the same manner as that Hinterleitner *et al.* reported. Thin films are deposited in different conditions of base pressure of the sputtering system and on different substrates (*n*, *p* and undoped Si). The large value of $S \sim -1098 \pm 100$ µVK$^{-1}$ was observed at low temperatures near 300 K in the Fe-V-W-Al base films of amorphous structure deposited on *n*-Si substrate. The power factor (*PF*) and figure of merit (*ZT*) are calculated to be $\sim 33.9$ mWm$^{-1}$K$^{-2}$ and 3.9, respectively. These values are highest among the values ever reported till date in Fe-V-W-Al based thin films. This observed anomalous behaviour has been attributed to crystallization of thin films in amorphous structure and the composite effect of thin film and substrate.

## II. EXPERIMENTAL DETAILS

**Sample synthesis:** Fe-V-W-Al based thin films were prepared using the RF magnetron sputtering (commercial VTR-150M/SRF RF magnetron sputtering system- ULVAC Kiko, Japan) using a single circular target of stoichiometrically prepared Fe$_2$VAl alloy ($\sim 2$ inches in diameter) and Fe/W/Al elemental chips. The distance between the target and substrate was $d = 60$ mm. The initial base pressure of the system was kept 10$^{-4}$ Pa. Argon gas with a pressure $\sim 2$ Pa was used for the deposition of all the samples presented in the paper. We have used three different substrates- *n*, *p* and undoped silicon wafer for synthesizing the thin films, the lattice parameters of which fit very well with those of Fe$_2$VAl. The details of the substrates are given in supplementary information (Supplementary Note 1). The base pressure of the sputtering system was varied from 0.1 $\sim 1.0$ X 10$^{-2}$ Pa for the deposition of thin films. The deposition temperature of the substrate was kept at 1050 K for all samples. The prepared samples are not annealed, rather vacuum quenched to room temperature after deposition. The deposition rate of the alloy was kept $\sim 23$ nm/min to prepare the thin films.

**Characterization techniques**: The crystal structure and phase purity of the prepared samples were confirmed using the X-ray diffraction (XRD) measurements (both in Bragg-Brentano and Grazing incident angle geometry) with conventional Cu-kα radiation ($\lambda = 1.54$ Å) equipped in the Bruker D8 Advance. A scanning electron microscope (SEM, Hitachi SU-6600) equipped with energy dispersive x-ray spectroscopy (EDX, JEOL JED-

2140 GS) was used to analyse the microstructure in the sample. The electrical resistivity as a function of temperature was measured in the temperature range 300 to 450 K under vacuum conditions in parallel to the film using a conventional DC four probe method. The temperature dependent Seebeck coefficient was estimated from the slope of $V$-$\Delta T$ plots using the steady state method with varying temperature difference in the range of 0 K $\leq \Delta T \leq$ 5 K. Both electrical resistivity and Seebeck coefficient were measured using homemade setups. Standard sample (constantan) was measured done before and after the measurement of all the thin films to evaluate the standard deviation in the measurement. The standard deviation in the Seebeck coefficient and electrical resistivity is found to be less than ±10% and ± 6%, respectively. The thermal conductivity of the thin film samples was measured at room temperature by use of the thermo-reflectance method along the perpendicular direction with the PicoTR developed by PicoTherm (Japan). A molybdenum film of 100 nm in thickness deposited on the sample was used as a transducer for the thermal diffusivity measurement in the PicoTR.

## III. RESULTS

### 1. X-ray diffraction: Variation in base pressure

**(i) θ-2θ scan:** Fig. 1 shows the room temperature XRD patterns (obtained in Bragg Brentano geometry) of Fe-V-W-Al thin films (~ 230 nm) deposited on *n*-type Si substrate at different base pressure of the sputtering machine (1) Sample 1- 0.1 X $10^{-2}$ Pa (2) Sample 2- 1.0 X $10^{-2}$ Pa. In the second case, oxygen was intentionally introduced to increase the base pressure of the sputtering system. Both XRD patterns include contribution from *n*-type silicon substrates (i.e. fundamental peaks Si (200) and Si (400)). In case of sample 1, presence of sharp fundamental peak (220) is noted whereas superlattice reflections (111) and (200) are absent. This indicate towards the formation of crystalline A2 Heusler structure (body-centred cubic (bcc), space group *Im-3m*) in this sample, which is in analogy with the crystal structure of $Fe_2V_{0.8}W_{0.2}Al$ thin films reported by Hinterleitner *et al.* [22]. Contrary to this, in sample 2, broadening of fundamental peak (200) is observed, which signifies the formation of amorphous phase rather than forming A2 (chemically disordered bcc structure), B2 (CsCl structure), or $L2_1$ structure (Heusler structure).

**(ii) Grazing incident angle XRD:** In addition to the XRD with Bragg Brentano geometry, XRD patterns of both samples are also obtained from grazing incident angle XRD to precisely extract the crystal structure of deposited Fe-V-W-Al thin film. The obtained patterns are plotted in Fig. 2. During this measurement, the incident angle was fixed at 1°. Presence of (220) and (400) peaks was confirmed in sample 1, whereas only a broader, diffuse peak is noted near 44.5° in sample 2. This further confirms the formation of crystalline Heusler phase and amorphous phase in sample 1 and 2, respectively.

**(iii) Electron diffraction (ED) pattern**: To derive the detailed information about the crystal structure of sample 1 and 2, electron diffraction (ED) patterns are obtained. Fig. 3 (a) depicts rings constructed by the discrete dots for sample 1, which reflects the polycrystalline nature of the thin film. We have estimated the $d$-spacing corresponding to each circle and corresponding reflections are identified (given in Table 1). From the $d$-spacing, the lattice parameter of sample 1 is determined to be ~ 2.969 Å, which is analogy with that of previous reported $Fe_2VAl$-based Heusler phases [22]. However, in case of sample 2 (Fig. 3 (b)), diffused rings are observed, which further validates the formation of amorphous structure in the sample. We have discussed the XRD patterns obtained for samples deposited on different substrates under different base pressure in the supplementary information (Note 2).

From the above observations, it can be concluded that change in base pressure of the sputtering system can significantly alter the phase of the Fe-V-W-Al thin films irrespective of nature of substrate of deposited thin film.

## 2. Surface morphology and compositional analysis:

Significant variation in the crystal structure was observed by XRD and TEM measurements for the deposited Fe-V-W-Al thin films. In supplement to new picture of structural evolution in our deposited samples (samples 1 and 2), compositional analysis and depth profiling would help to view deeper into chemistry of the crystalline and amorphous phase, which is described below:

**Sample 1:** In Fig. 4 (a), a cross sectional scanning transmission electron microscopy (STEM) image of the thin film-interface-substrate obtained at 200 nm scale for sample 1 is shown. A diffusion zone is noted between thin film and substrate, which extends into Si substrate. This interlayer is not compact in size, rather appears as weakly connected small islands. In order to know the information about the type of elemental diffusion into substrate, cross sectional STEM image observation and EDX mapping are also performed in the range of 20 nm scale (shown in Fig. 4 (b)- (j)). Significant diffusion of element into substrate, which can be attributed to Fe and V diffusion, is clearly observed for crystalline sample 1. However, it is observed that this diffusion is not uniform throughout the interface, rather there is presence of small islands. All the elements are uniformly distributed throughout the sample.

In this crystalline sample, no secondary phase is precipitated. Hence, to ascertain the elemental distribution throughout the sample and at the interface depth profiling across the cross-sectional area (obtained at 200 nm scale) is carried out. Fig. 5 depicts the depth profiling of elements across the cross-sectional area (200 nm scale) of the sample. From the fig. 5, thickness of thin film is determined to be ~ 286 nm, which is comparatively more than estimated based on sputtering rate. Whereas interface is estimated to be ~ 20 nm. At the interface, a clear diffusion of all elements (Fe, V, W, and Al) in Si substrate and vice versa is noted. Surprisingly, in spite of high vacuum conditions during sample preparation, we also observed the presence of oxygen (approx. 5 %) across the sample. From the depth profile, the average composition of the thin film is determined to be ~$Fe_2V_{0.763}W_{0.168}Al_{0.716}$, which slightly deviates from the composition of thin film $Fe_2V_{0.8}W_{0.2}Al$ reported by Hinterleitner *et al*.

**Sample 2:** Fig. 6 (a) illustrates the cross-sectional STEM image obtained for sample 2 at 200 nm scale. Similar to sample 1, a negligible diffusion layer is noticeable between thin film and substrate. In order to throw some light on the diffusion layer, cross sectional STEM image and EDX mapping is also done at 20 nm scale (shown in Fig. 6 (b)-(j)). Surprisingly, the diffusion layer contains only some small traces of Fe, which is comparatively negligible with respect to sample 1. Additionally, uniform distribution of elements throughout the sample is noted, which rules out the possibility of formation of secondary phase. Depth profiling across the cross-sectional area (~ 200 nm) is done to estimate the average composition of the thin film (shown in Fig. 7). From this analysis, the thickness of thin film is estimated to be ~ 326 nm, which is more compared to sample 1. This reflects that introduction of oxygen during the sample preparation can alter the thickness of the deposited thin film. We have also estimated the average oxygen content in the sample 2, which is 16-17 %. This value is quite higher than that noted in sample 1. The average composition of sample 2 is found to be ~ $(Fe_2V_{0.810}W_{0.184}Al_{0.859})_{0.84}O_{0.16}$, which reflects that sample gets oxidized in the presence of oxygen during sample preparation.

Interestingly, microcracks or voids are not visible in STEM images of both sample 1 and sample 2, which is noted in similar samples prepared at the substrate temperature of 743 and 843 K [25]. AFM images measured for the samples 1 and 2 deposited at 1050 K are shown in Fig. 8. The surface roughness of the thin film sensitively varied with the oxygen content during the deposition. The surface roughness ($R_a$) was ~1.44 nm for sample 2, where amorphous structure was developed. On the other hand, $R_a$ suddenly increased with lowering the oxygen concentration. i.e. for sample 1; $R_a$ ~ 2.88 nm. In the latter case, the large value of $R_a$ can be closely related to growth of crystalline nano-grains.

The above analysis signifies that introduction of oxygen can significantly influence the thickness of the thin film, surface morphology as well as the composition of the thin film. As noted from above, there is diffusion of elements into Si substrate in crystalline sample (the sample #1), whereas no significant traces of elements were noted in amorphous sample (the sample #2) at the interface.

## 3. Thermoelectric properties

Temperature dependence of electrical resistivity ($\rho$) in the temperature range 300-400 K is shown in Fig. 9 (a) for the sample #1 and the sample #2. At 300 K, the value of $\rho$ of the sample #1 is noted to be ~ 1.12 mΩcm, which is comparable with that reported for bulk $Fe_2VAl$ alloys [20, 26]. In sample 1, it is observed that the electrical resistivity moderately decreases with increasing temperature. This behaviour is generally exhibited by hopping conduction in disordered metals in the strong scattering limit known as the Mott-Ioffe-Regel limit [27]. In this particular case, electrons are strongly affected by presence of structural distortions (in form of atomic voids), anti-site disorder or strong electron-electron scatterings. The carrier localization is further enhanced for the system of small density of states in the vicinity of chemical potential, and -$Fe_2VAl$ based alloys possess this characteristic because of the presence of a pseudo-gap, i.e., of a deep minimum in the density of electronic states near the Fermi energy, $E_F$ [28, 29]. This type of behaviour in electrical resistivity was earlier reported in bulk $Fe_2VAl$ based Heusler alloys and $Fe_2VAl_{0.95}Si_{0.05}$ based thin films at high temperatures [1, 20, 26].

An unusual behaviour of $\rho$ is seen in the sample #2. At 300 K, the value of $\rho$ of sample 2 is noted to be ~ 4.30 mΩ-cm, which is quite higher than crystalline sample 1. In this sample, initially $\rho$ decreases with increase in temperature till 320 K. But, beyond 320 K, it exhibits positive temperature coefficient (PTC) till 400 K. This behaviour is quite different from that previously reported for $Fe_2VAl$ based thin films [1, 30, 31, 32]. Notably, PTC was seen in bulk Heusler alloys $Fe_2VAl_{1-x}Si_x$ and $Fe_2VAl_{1-x}Sn_x$ in low temperature regime (100- 300 K) [33]. Whereas no suitable explanation is given in literature to explain this anomalous behaviour.

Fig. 9 (b) shows the temperature dependence of Seebeck coefficient $S(T)$ measured for the samples #1 and #2 in the temperature range 320- 400 K. Both samples exhibit negative values in the measured temperature range. This indicates that majority charge carriers are electrons. Similarly, n-type behaviour was previously reported in the bulk and thin film of $Fe_2VAl$ based Heusler alloys.

For sample #1, with increasing temperature, $S$ shows initial increase in magnitude up to 340 K, followed by decrease. The maximum value of $|S|$ is noted ~ 88 ± 10 μVK$^{-1}$ near 340 K. This magnitude is slightly lower than previously reported value in bulk samples with $L2_1$ structure, which is ~ –120 - 150 μVK$^{-1}$ and $Fe_2VAl_{0.95}Si_{0.05}$ based thin film samples [20, 26]. The lower value of $|S|$ is presumably attributed to the presence of anti-site disorder which can alter the shape of pseudogap near $E_F$ in the electronic density of states. For the sample #2, the maximum value of $|S|$ is noted ~ 1098 ± 100 μVK$^{-1}$ near room temperature. This value is also exceptionally larger than previously reported values of |S| in Fe$_2$VAl based thin films. The maximum value of |S| ever reported for Fe$_2$VAl based materials is approximately 500 µV/K in Fe$_2$V$_{0.8}$W$_{0.2}$Al thin films by Hinterleitner *et al*. [22].

We also calculated the power factor ($PF$) = $S^2/\rho$ for each prepared sample. The obtained temperature dependence of power factor is plotted in Fig. 9 (c). A large power factor i.e. 33.9 mWm$^{-1}$K$^{-2}$ is obtained near room temperature in case of the sample #2, which is about 10 times larger than obtained for Bi-Te based thermoelectric materials and is smaller than that reported for Fe$_2$V$_{0.8}$W$_{0.2}$Al thin films [22].

Thermal conductivity $\kappa$ of the sample #1 and the sample #2 is obtained at 320 K by means of time domain thermo-reflectance (TDTR) method with rear heating and front detection configuration. Figs. 10 (a)- (b) represents the typical transient temperature curve of TDTR obtained at $T$ = 320 K for the sample #1 and the sample #2. The solid red line depicts the theoretical least square fitting of the TDTR signal of the front face heating from face measurement configuration. For sample 1, the value of $\kappa$ is estimated to be ~ 3.90 Wm$^{-1}$K$^{-1}$. This obtained value is similar to that obtained in W substituted Fe$_2$VAl bulk Heusler alloy ($\kappa$ ~ 5.0 Wm$^{-1}$K$^{-1}$) [20] and Fe$_2$VAl based thin films [34]. The lower value in prepared thin film (sample 1) can be ascribed to phonon scattering by point defects resulting from the presence of heavier W atoms. Interestingly, in sample #2, $\kappa$ is found to be ~ 2.75 W/m-K which is reduced value in comparison with sample 1. The lower value of $\kappa$ in sample #2 can be attributed to presence of amorphous structure which promotes higher electron-phonon scatterings.

The dimensionless figure of merit $ZT$ for both samples were also evaluated near 320 K. In sample #2, the low value of $\kappa$ and large power factor, results in high value of figure of merit ($ZT$) ~ 3.9. The obtained value is the higher value among the ones ever reported in literatures. Among recently reported large values of figure of merit are $ZT$ ~ 6 in case of Fe$_2$V$_{0.8}$W$_{0.2}$Al thin films [22], $ZT$ ~ 2.4 in artificial layers of Bi$_2$Te$_3$ and Sb$_2$Te$_3$ [35] near room temperature; $ZT$ ~ 2.6 in single crystalline [36, 37], SnSe; $ZT$ ~ 3.1 in hole doped polycrystalline SnSe [38] and $ZT$ ~ 2.5 in $p$ type PdTe-SrTe [39].

## IV. DISCUSSION

**(a) Effect of substrate and crystal structure -** As noted above, an anomalously large value of $S$ is noted in oxidized amorphous sample (sample #2) as compared to crystalline sample (sample #1). This implies an important role of crystal structure in deciding the magnitude of $S$ of the prepared samples. To get a closer look and understanding of individual contributions of the active thin film layer (Fe-V-W-Al), the interlayer and the substrates, we have also synthesized samples on different substrates- $p$ type Si (sample #3 and #5) and undoped Si (sample #4 and #6). The crystal structure of the prepared samples has been characterized using the XRD and the details about the crystal structure is given in the supplementary information Note 2. Thermoelectric behaviour of the prepared samples is discussed below.

**(i) $p$ type Si substrate-**

The samples #3 and #5 are prepared in similar manner as sample #1 and #2, respectively. The composition of these samples was found to be similar to sample #3 and sample #5. As noted from Fig. S1, sample #3 and sample #5 crystallizes in crystalline A2 and amorphous structure, respectively. Electrical resistivity $\rho$ and Seebeck coefficient $S$ were measured for both samples. Fig. 11 (a) represents $\rho(T)$ of for samples 3 and 5 plotted as function of temperature in the range 300- 370 K. In case of sample #3, $\rho_{300 K}$ is ~ 1.05 mΩ-cm, which is quite smaller than noted for sample #1 (deposited in $n$ type Si wafer). The magnitude of $\rho$ moderately increases with increasing temperature. The observed behaviour is similar to typical $\rho$ behaviour noted in metallic systems of ordered structure but quite different from the trend observed in sample #1, where negative temperature coefficient of electrical resistivity (TCR) is observed.

For sample #5, $\rho_{300 K}$ is ~ 6.76 mΩ cm, which is quite higher than noted for sample #2 (deposited on $n$ type Si wafer). In this sample, positive TCR as that in metals of an ordered crystal structure is observed throughout the measured temperature range, which is different from that in other Fe-V-W-Al thin films.

Fig. 11(b) represents the temperature dependent $S$ of crystalline sample #3 and amorphous sample #5. Surprisingly, both samples show positive values over the whole measured temperature range from 300 to 370 K. This fact reflects that majority charge carriers are holes in both samples. In the crystalline sample #3, $S$ initially increases with temperature until 340 K, followed by decrease with increasing temperature. In this case, the maximum value of the $S$ is noted to be ~ 118 ± 12 µVK$^{-1}$ near 320 K. However, for the amorphous sample #5, the maximum value of |$S$| is greatly enhanced to ~ 1295 ± 100 µVK$^{-1}$ near 310 K. This value is also exceptionally larger than previously reported values of $S$ in p-type Fe$_2$VAl based thin films and bulks. In both cases, it can be noted that values of $S$ are positive, which is contrast to previously reported values of $S$ in thin films, bulk materials [20, 22, 34] and investigated samples #1 and #2.

**(b) undoped Si substrate-**

Similar to previously prepared samples, Fig. S2 confirms that the sample #4 and sample #6 prepared w/wo residual oxygen possess crystalline and amorphous structure, respectively. The composition of these samples was also confirmed to be similar to sample #3 and sample #5. For both samples, $\rho$ and $S$ has been measured in the same manner as those of previous samples. Temperature response of $\rho$ for samples #4 and #6 in the measured temperature range is shown in Fig. 12 (a). For the crystalline sample #4 deposited on non-doped Si substrate, $\rho_{300\,K}$ is ~ 0.62 mΩ-cm, which is comparable with that of crystalline sample #3 on p-type Si substrate. In the sample #4, it is observed that the electrical resistivity moderately decreases with increasing temperature. This behaviour is similar to seen in sample #1, which exhibits negative TCR. In the case of sample #6, a larger value of $\rho_{300\,K}$ ~ 57.86 mΩ-cm is observed. This value is quite higher than reported in bulk and thin film Fe$_2$VAl samples. It moderately decreases with increasing temperature. Less enhanced negative TCR is often observable in heavily disordered metals, and the small magnitude would be related with the small density of states near the chemical potential. The reason for the absence of localization tendency is still unrevealed.

Fig. 12 (b) depicts the temperature dependence of $S$ observed for the samples #4 and #6. Both samples exhibit negative $S$ over the whole measured temperature range. This reflects that majority charge carriers are electrons in both samples. In the sample #4, the magnitude of $S$ monotonically and moderately increases with increasing temperature such as that in typical metals. The maximum value of the observed |S| is ~ 35.0 ± 3.5 µVK$^{-1}$ near 400 K. This magnitude is lower than that expected for *n*-type Fe$_2$VAl based thin films and bulks. In the sample #6, |S| drastically decreases with increasing temperature at 300 K < *T* < 355 K but shows a significant increase with temperature at above 360 K. The maximum value of |S| is found to be ~ 120 ± 15 µVK$^{-1}$ near 310 K and 400 K. This magnitude is higher than that of the sample #4 but similar to those reported in bulk Fe$_2$VAl alloys.

The findings described above suggest that crystallinity of the structure and nature of substrate are two crucial factors to observe unexpectedly high values of |S| in Fe-V-W-Al based thin films. As noted from above observations, the amorphous samples #2, #5 and #6 exhibit high value of |S| (i.e. the sample #2 possesses *S* = –1098 ± 100 µVK$^{-1}$ ; the sample #5 ~ 1250 ± 100 µVK$^{-1}$; the sample #6 ~ –120 ± 15 µVK$^{-1}$) with respective to crystalline counterparts (the sample #1 ~ –88 ± 10 µV/K; the sample #3 ~ 118 ± 12 µVK$^{-1}$; the sample #4 ~ –35.0 ± 3.5 µVK$^{-1}$ near 350 K). This reflects the importance of formation of amorphous oxide under the presence of residual oxygen in atmosphere of sputtering process.

Amorphous structure naturally results in the observation of high *ρ* such as that in the sample #2, #5 and #6. However, the values observed for these samples are much smaller than that of Si substrates, and therefore the charge carriers are naturally considered to pass through the deposited thin film only. Hence the behaviour in *ρ*(*T*) observed for these samples is dominated by the oxide amorphous phase, and the slightly high values of *ρ* is attributed to presence of high amount of oxygen which plays important role in the formation of amorphous phase.

Interestingly, the nature of substrate greatly influences the magnitude and sign of the observed *S*. In the samples deposited on *n*-type Si substrate, negative values of *S* with large magnitude are observed, whereas in samples deposited on *p* type Si substrate show large positive values. Although the magnitude is rather small, the same tendency in the sign of *S* is observable for the crystalline samples. This means that the sign of the Seebeck coefficient is determined by the nature of substrate, and the large magnitude will be related with the formation of amorphous oxide layer.

Recently reported thermoelectric behaviour of $Fe_2V_{0.8}W_{0.2}Al$, the high value of $S$, was ascribed to two factors (1) the formation of chemically disordered bcc structure (A2-phaes) in prepared thin film and (2) large change in slope of electronic density of states in the vicinity of Fermi level [22]. The formation of disordered bcc structure was attributed to the rather small lattice thermal conductivity and modification in electronic structure near the chemical potential. The latter are capable of enhancing the Seebeck coefficient and thermoelectric figure of merit, despite that the mechanism leading to such a characteristic in electronic structure near the chemical potential is still controversial issue. It should be emphasized that, in the present study, we have definitely proved that formation of chemically disordered bcc structure (A2 phase) solely does not significantly affect the thermoelectric properties as is shown in the sample #1, #3, and #4. In these samples, we have observed a small value of $S$, which also depends on the nature of the substrate. Generally, the electrical filed is well screened by mobile carriers in a very short length range. Therefore, it is rather difficult to understand the observation of large thermoelectric voltage generated by substrate through the metallic Fe-V-W-Al-O amorphous layer. Indeed, our preliminary performed finite element analyses certainly indicate the difficulty in reproducing the observation of large Seebeck coefficient, that is generated from the substrate, across a metallic layer.

**2) Contribution from interlayer –**

As discussed in section 2 "Surface morphology and compositional analysis", there is significant diffusion of Si substrate into Fe-V-W-Al thin film and Fe, V into Si substrate in the case of sample #1 possessing small magnitude of $S$, whereas no significant diffusion of any element in Si substrate is seen in the sample #2 possessing the large magnitude of $S$. This implies that interlayer does not significantly contribute towards higher value of $S$ in amorphous samples. Thus, we have neglected the diffusion of Si into Fe-V-W-Al thin film and vice versa as one of the main factors for observed large value of $S$ in this study. However, the diffusion of Si into thin film can contribute towards lower value of $\rho$ as noted in the samples #1, #3 and #4.

### 3) Contribution from formation of secondary phases-

In the studied samples,-the formation of cracks or the precipitation of secondary phase were not observed on the surface of all the investigated samples. This fact signifies that secondary phase has negligible impact on the observed behaviour of the Fe-V-W-Al thin films.

Higher surface roughness can also result in increment in $\rho$ [40], thus, larger magnitude of $\rho$ is naturally expected in the sample #1 than that of-the sample #2. However, sample #1 shows much reduced value, thus implying less significant contribution from the surface roughness towards the observed $\rho$.

### 4) Phonon and magnon drag effect-

Non-equilibrium phonons are generated in the substrate due to applied thermal gradient. Part of these phonons imparts their momentum to electrons resulting in electrical current. Hence, an extra electric field is established which counters the generated flow of electrons. The contribution from phonon drag becomes significant only in the temperature range where electron- phonon scattering is most pronounced [40, 41]. At very low temperatures, number of phonons that can contribute the momentum transfer becomes much negligibly small, so that the phonon drag effect becomes also negligible. At high temperatures above the Debye temperature, where phonon-phonon scattering is dominant, we observe almost no phonon drag effect neither. The electron-phonon inelastic scattering becomes most dominant near the temperature range where the peak of lattice thermal conductivity is observable, and it is generally much lower than room temperature. Therefore, the extremely large Seebeck coefficient observable at room temperature is supposed not to related with the phonon-drag effect.

Magnon drag effect is the magnetic analogue of the phonon drag effect. In this effect, applied thermal force is exerted on magnons via phonon-magnon coupling i.e. the magnon flux drags the electron flux via s-d interactions [42]. This type of effect is usually noted in ferromagnetic materials. In these cases, the observed thermopower is usually 1 to 2 orders higher in magnitude in comparison to diffusion thermopower. Recently, it was reported that the large Seebeck coefficient and power factor observed in $Fe_2V_{0.8}W_{0.2}Al$ thin films is likely due to a magnon drag effect related to the tungsten-based impurity band [23]. However, in the present investigated samples, magnon drag effect cannot be explain the behaviour of huge *S* coefficient observed in amorphous samples. In the present case, tungsten-based impurity bands are supposed to present in all the prepared samples, however, high value of *S* is noted only in amorphous samples.

From the above discussions, it can be concluded that both oxidized amorphous structure and composite effect of thin film and substrate plays a crucial role in deciding the magnitude of *S* and *PF* in the Fe-V-W-Al thin films.

## V. CONCLUSIONS

In conclusion, electrical resistivity, Seebeck coefficient and thermal conductivity measurements has confirmed the presence of substantially high value of figure of merit (*ZT* ~ 3.9) in Fe-V-W-Al non Heusler phase thin film deposited on *n* type Si substrate. Systematic analysis of room temperature XRD and electron diffraction reveals the formation of crystalline A2 Heusler structure in oxygen deficient samples, which transforms into oxidized amorphous phase on increment of oxygen content in thin films. In addition to this, our substrate dependent study also confirms that nature of substrate (*p*, *n* and undoped Si) plays a key role in observing anomalous behaviour in Fe-V-W-Al based thin films.

## VI. SUPPORTING INFORMATION

The following Supporting Information is available free of charge at the website.

1. Supplementary note 1: Information about the substrate used in the present study.
2. Supplementary note 2: Analysis of crystal structure of samples prepared by variation of base pressure and substrate.
3. Figure S1- Room temperature XRD patterns of sample #3 and sample #5 obtained in Bragg-Brentano geometry.
4. Figure S2- Room temperature XRD patterns of sample #4 and sample #6 obtained in Bragg-Brentano geometry.

**Tables**

Table 1 *d* spacing and corresponding (hkl) planes obtained from electron diffraction pattern of sample #1

| *d* spacing | Intensity | h | k | L |
| --- | --- | --- | --- | --- |
| 2.10 | 100 | 1 | 1 | 0 |
| 1.47 | 12.3 | 2 | 0 | 0 |
| 1.20 | 19.1 | 2 | 1 | 1 |

# Figures

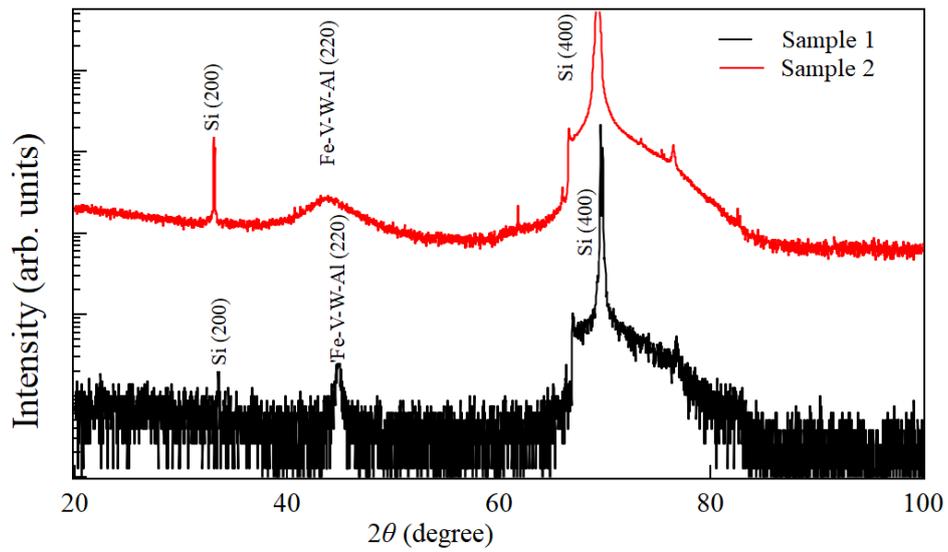

**Figure 1** Room temperature XRD patterns of sample #1 and sample #2 obtained in Bragg-Brentano geometry.

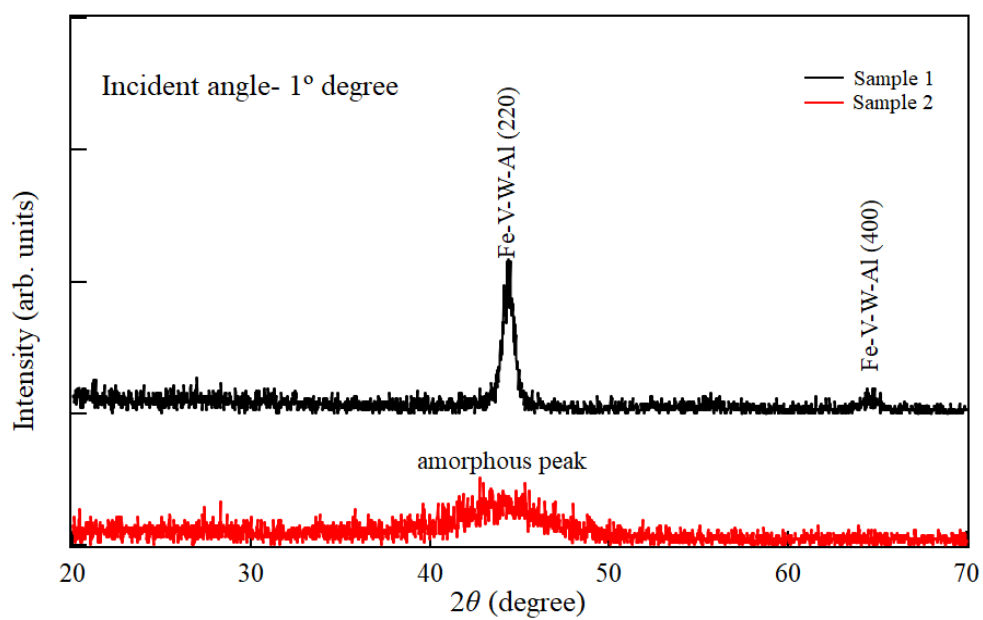

**Figure 2** GIXRD patterns of sample #1 and sample #2 obtained at fixed incident angle ~ 1°.

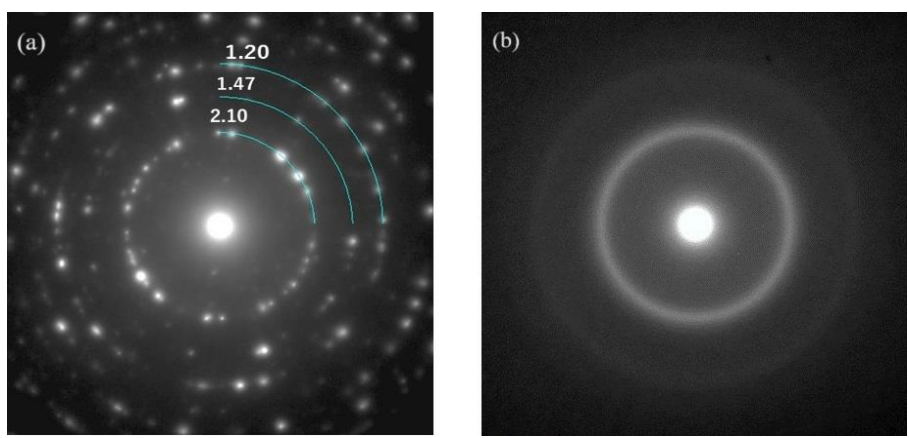

**Figure 3** Electron diffraction patterns for (a) sample #1, green line represents the circles indicating different $d$ spacing. (b) sample #2

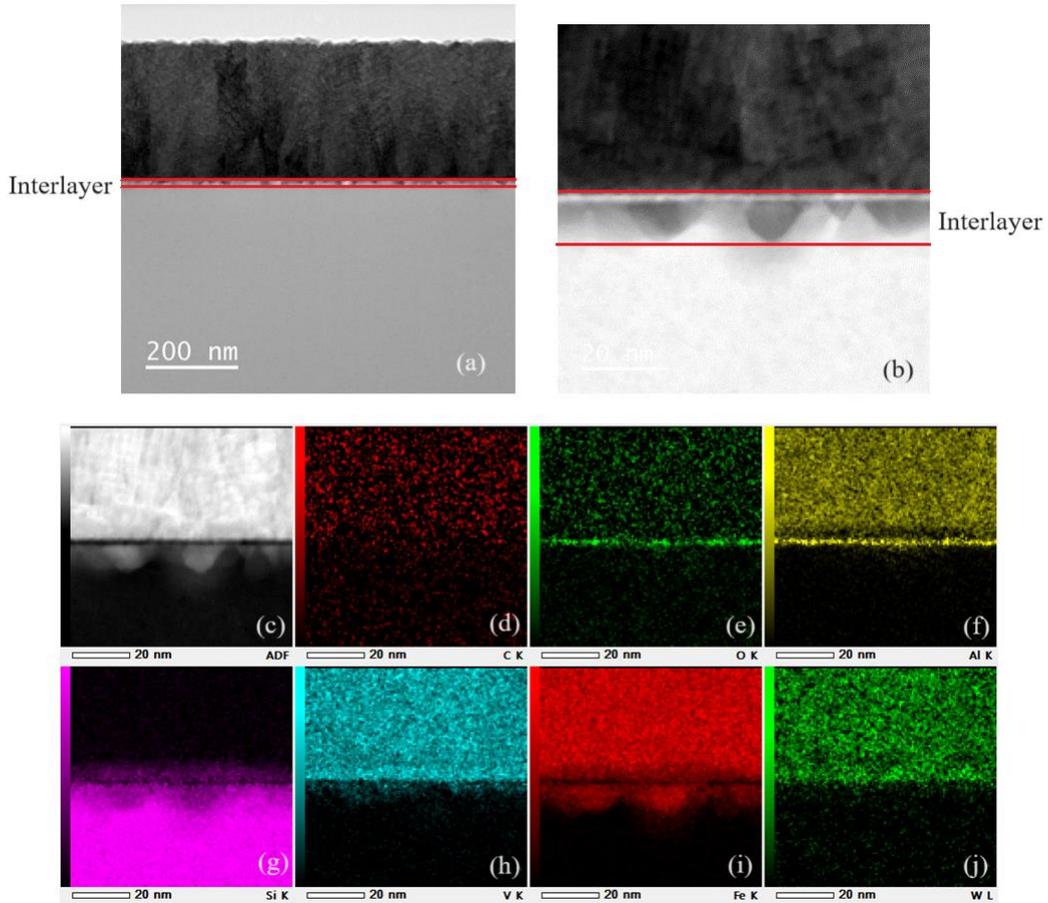

**Figure 4** Cross sectional STEM image of sample #1 at (a) 200 nm scale (b) 20 nm scale. Red lines denote the interlayer between thin film and substrate (c) Annual dark field cross sectional image of the sample #1. TEM-EDX mapping of (d) Carbon (e) Oxygen (f) Aluminium (g) Silicon (h) Vanadium (i) Iron (j) Tungsten.

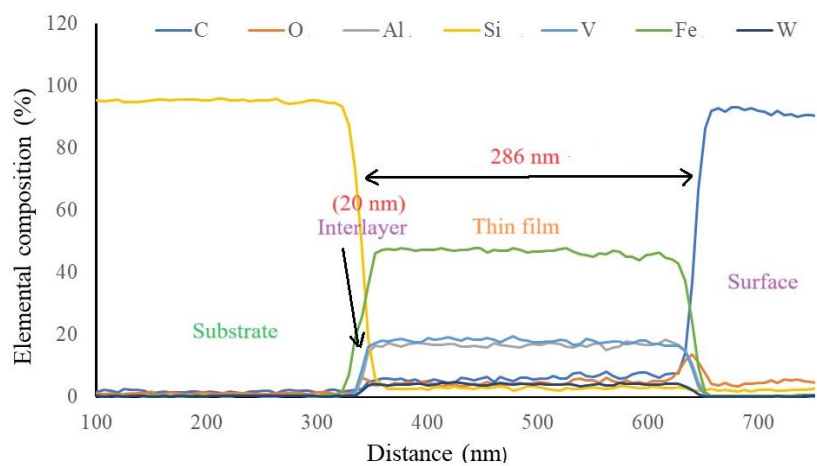

**Figure 5** Elemental composition (%) as a function of distance (nm) from the surface for sample #1.

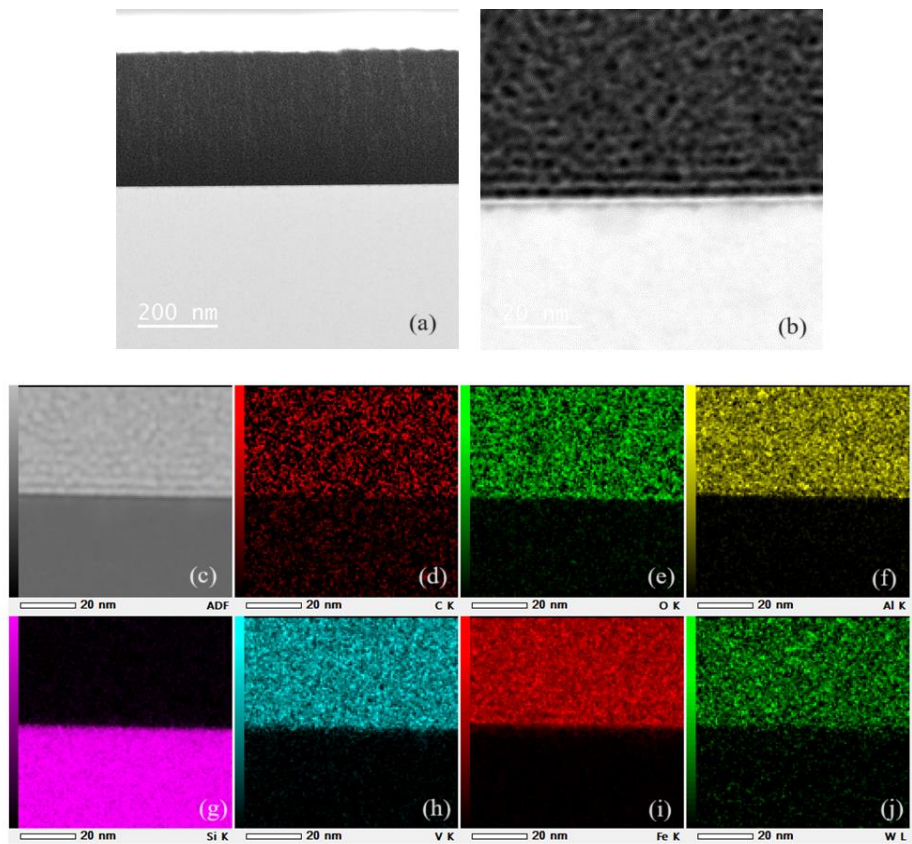

**Figure 6** Cross sectional STEM image of sample #2 at (a) 200 nm scale (b) 20 nm scale. Red lines denote the interlayer between thin film and substrate (c) Annual dark field cross sectional image of the sample #2. TEM-EDX mapping of (d) carbon (e) oxygen (f) aluminium (g) silicon (h) vanadium (i) iron (j) tungsten.

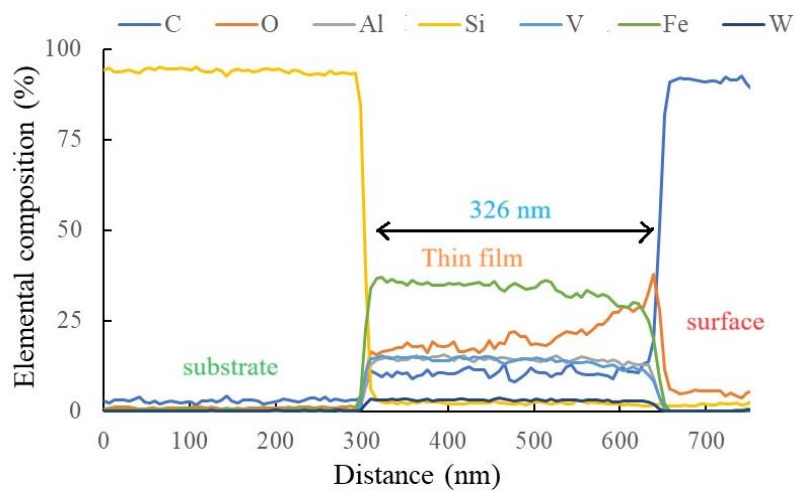

**Figure 7** Elemental composition (%) as a function of distance (nm) from the surface for sample #2.

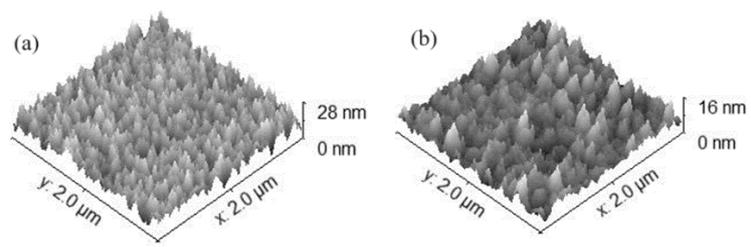

**Figure 8** 3D AFM images of (a) sample #1 and (b) sample #2.

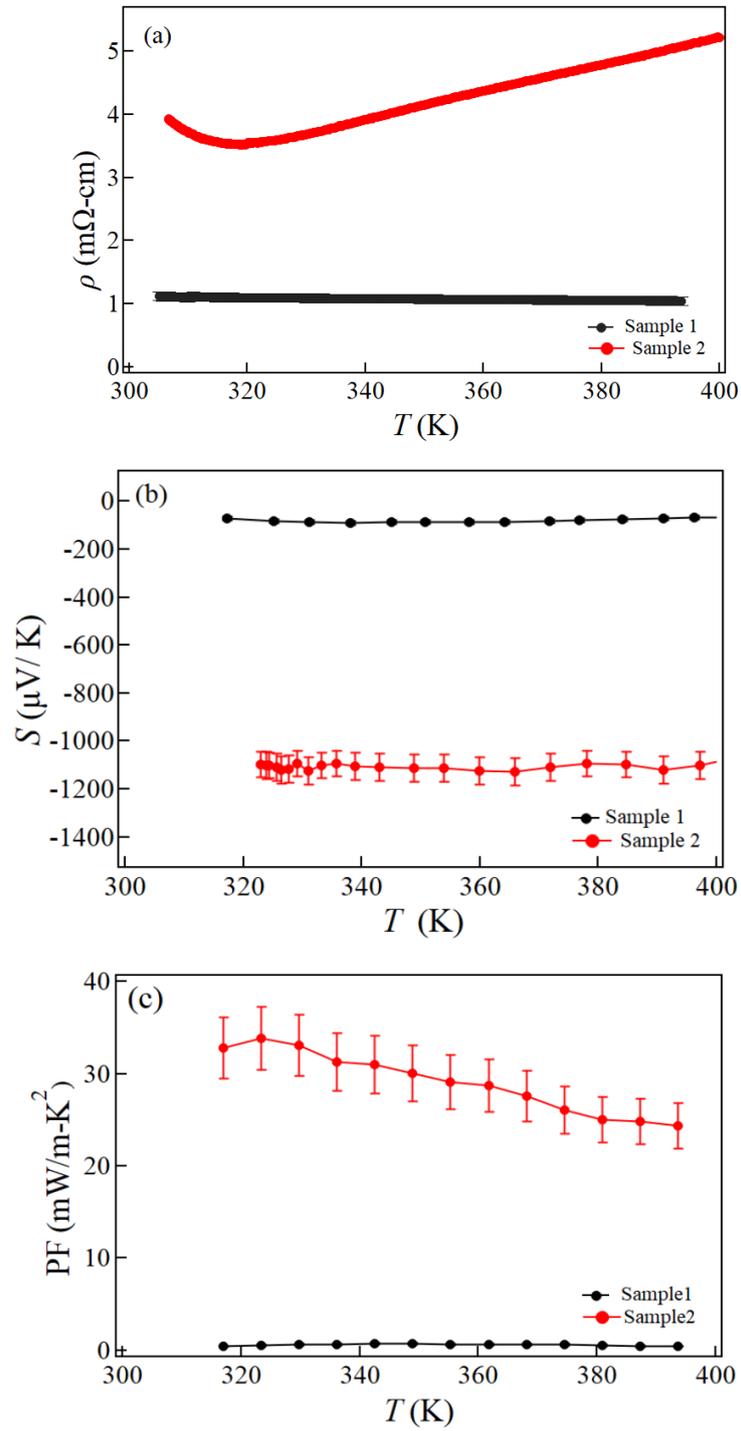

**Figure 9** Temperature dependence of (a) Electrical resistivity (b) Seebeck coefficient (c) Power factor measured in the temperature range 300- 400 K for sample #1 and sample #2.

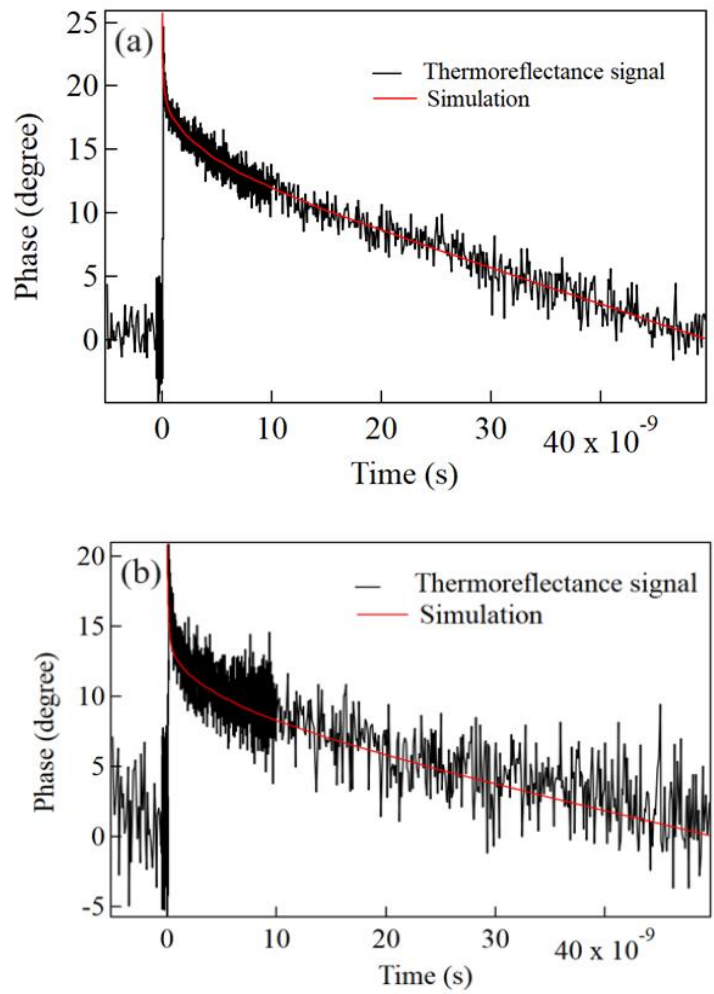

**Figure 10** Time-dependent temperature response curve of (a) sample #1 and (b) sample #2.

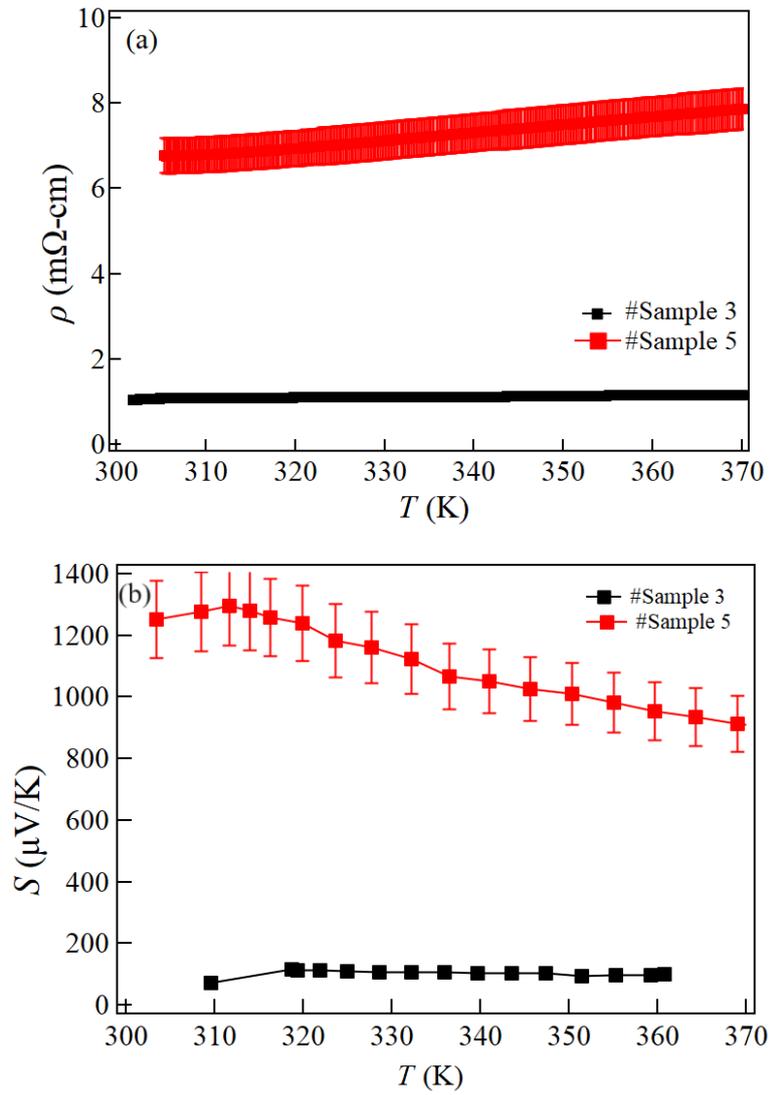

**Figure 11** Temperature dependence of (a) Electrical resistivity (b) Seebeck coefficient measured in the temperature range 310- 370 K for sample #3 and sample #5.

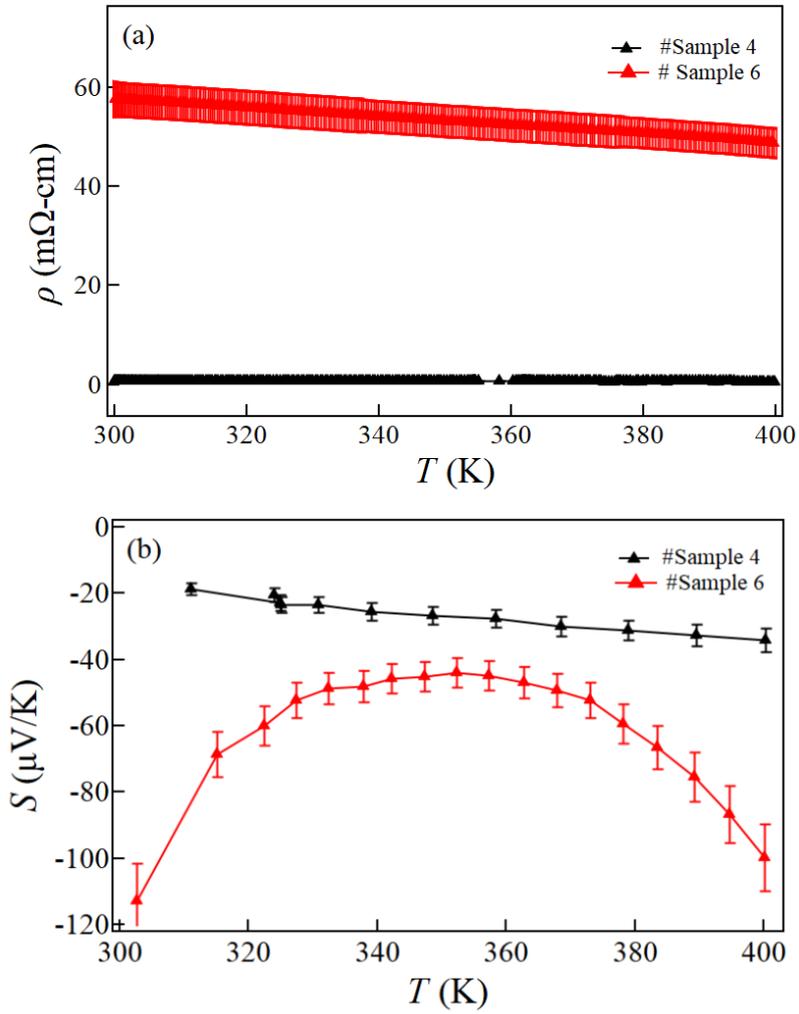

**Figure 12** Temperature dependence of (a) Electrical resistivity (b) Seebeck coefficient (in the temperature range 300- 400 K for sample #4 and sample #6.